\documentstyle[11pt,paspconf,psfig]{article}

\def\edcomment#1{\iffalse\marginpar{\raggedright\sl#1\/}\else\relax\fi}
\reversemarginpar

\def\kms{{\rm\,km\,s^{-1}}}

\def\msun{{\,M_\odot}}
\def\cm{{\rm\,cm}}

\def\frac#1#2{{#1 \over #2}}
\newbox\grsign \setbox\grsign=\hbox{$>$} \newdimen\grdimen \grdimen=\ht\grsign
\newbox\simlessbox \newbox\simgreatbox
\setbox\simgreatbox=\hbox{\raise.5ex\hbox{$>$}\llap
     {\lower.5ex\hbox{$\sim$}}}\ht1=\grdimen\dp1=0pt
\setbox\simlessbox=\hbox{\raise.5ex\hbox{$<$}\llap
     {\lower.5ex\hbox{$\sim$}}}\ht2=\grdimen\dp2=0pt
\def\simgt{\mathrel{\copy\simgreatbox}}
\def\simlt{\mathrel{\copy\simlessbox}}

\def\kms{{\rm km/s}}

\def\cm2{{\rm cm ^{-2}}}
\def\mum{\mu {\rm m}}

\def\deg{^\circ}

\def\wisk#1{\ifmmode{#1}\else{$#1$}\fi}

\def\kms{\ifmmode {\>{\rm km\ s}^{-1}}\else {km s$^{-1}$}\fi}
\def\gtrapprox{\;\lower 0.5ex\hbox{$\buildrel >
    \over \sim\ $}}             
\def\lessapprox{\;\lower 0.5ex\hbox{$\buildrel < \over \sim\ $}}

\def\psqcm{\ifmmode {\>{\rm cm}^{-2}}\else {cm$^{-2}$}\fi}
\def\pcubcm{\ifmmode {\>{\rm cm}^{-3}}\else {cm$^{-3}$}\fi}
\def\be{\begin{equation}}
\def\ee{\end{equation}}
\def\bea{\begin{eqnarray}}
\def\eea{\end{eqnarray}}
\def\rpcsq{\ifmmode {r_{\rm pc}^2}\else {$r_{\rm pc}^2$}\fi}
\def\rpc{\ifmmode {r_{\rm pc}}\else {$r_{\rm pc}$}\fi}
\def\fabs{\ifmmode {f_{\rm abs}}\else {$f_{\rm abs}$}\fi}
\def\msol{\ifmmode {\>M_\odot}\else {$M_\odot$}\fi}
\def\lsol{\ifmmode {\>L_\odot}\else {$L_\odot$}\fi}

\def\msolar{\ifmmode{M_\odot}
    \else{$M_\odot$}\fi}
\def\lsolar{\ifmmode{L_\odot}
    \else{$L_\odot$}\fi}
\def\msun{\ifmmode{M_\odot}
    \else{$M_\odot$}\fi}
\def\lsun{\ifmmode{L_\odot}
    \else{$L_\odot$}\fi}
\def\@versim#1#2{\lower 2.9truept \vbox{\baselineskip 0pt \lineskip 
    0.5truept \ialign{$\m@th#1\hfil##\hfil$\crcr#2\crcr\sim\crcr}}}
\catcode`\@=12

\def\ion#1#2{\ifmmode \mbox{{\rm #1}}\,\mbox{{\sc #2}} 
        \else {\rm #1}$\,${\sc #2}
        \fi}

\def\unitspace{\,}                      

\def\un#1{\ifmmode \unitspace{\rm #1} 
          \else $\unitspace$#1
          \fi}
\def\pun#1#2{\ifmmode \unitspace\mbox{\rm #1}^{#2} 
             \else $\unitspace$#1$^{#2}$
             \fi}

\def\kms{\un{km}\pun{s}{-1}}          
\def\Lsun{\ifmmode \un{L}_{\odot}     
          \else $\un{L}_{\odot}$
          \fi}
\def\mum{\ifmmode \unitspace\mu\mbox{\rm m} 
         \else $\unitspace\mu$m
         \fi}

\def\sou#1#2{\relax                       
             \ifmmode {\rm #1}\,{\rm #2}  
             \else #1$\,$#2
             \fi}

\def\simlt{\mathrel{\hbox{\rlap{\hbox{\lower4pt\hbox{$\sim$}}}\hbox{$<$}}}}
\def\simgt{\mathrel{\hbox{\rlap{\hbox{\lower4pt\hbox{$\sim$}}}\hbox{$>$}}}}
\def\simgreat{\mathrel{\copy\simgreatbox}}

\def\kms    {\ifmmode{{\rm km~s}^{-1}}\else{km~s$^{-1}$}\fi}
\def\hcccn     {HC$_3$N}
\def\hcop       {HCO$^{+}$}

\def\hhco       {H$_2$CO}
\def\hh         {H$_2$}
\def\hhcotcm    {$2_{11} - 2_{12}$}
\def\hhcoscm    {$1_{10} - 1_{11}$}
\def\hhcoparags {$1_{01} - 0_{00}$}

\def\pccm   {cm$^{-3}$}
\def\pscm   {cm$^{-2}$}
\def\hh         {H$_2$}
\def\mhd    {\ifmmode {n_{{\rm H}_2}} \else $n_{{\rm H}_2}$\fi}
 
\def\Tkin    {$T_{\rm kin}$} 
\def\Trot   {\ifmmode{T_{\rm rot}}\else$T_{\rm rot}$\fi}

\begin{document}

\title{Interferometric Observations of Redshifted Molecular Absorption
toward Gravitational Lenses}
\author{Karl M. Menten}
\affil{Max-Planck-Institut f{\"u}r Radioastronomie, Auf dem H{\"u}gel 69,
  D-53121 Bonn, Germany (kmenten@mpifr-bonn.mpg.de)}
\author{Christopher L. Carilli}
\affil{National Radio Astronomy Observatory, P.O. Box 0, Socorro, NM 87801,
  USA (ccarilli@aoc.nrao.edu)}
\author{Mark J. Reid}
\affil{Harvard-Smithsonian Center for Astrophysics, 60 Garden Street,
  Cambridge, MA 02138, USA (reid@cfa.harvard.edu)}

\begin{abstract}

We have made radio- and millimeter-wavelength interferometric
observations of a variety of molecular absorption lines toward the
gravitational lens systems B0218+357 and PKS~1830$-$211. The absorption
occurs in the lensing galaxies at redshifts of 0.685
and 0.89, respectively. The high spatial resolution of our
VLA and VLBA observations allows imaging of the background continuum
emission and the absorption distribution. Our multi-transition
studies yield estimates of the cosmic microwave background temperature at
$z = 0.89$ and determinations of  molecular and isotopic abundance ratios,
which allow meaningful comparisons with Galactic molecular clouds.

\end{abstract}

\index{HI 21cm Absorption}  \index{Quasar Absorption Lines}  
\index{Red Quasars}  
\index{Molecular Absorption}  \index{Microwave Background}  \index{HCO$^+$}  
\index{HCN}  \index{C$_3$H$_2$}  \index{HC$_3$N}  \index{VLBA}  \index{VLA}  
\index{Gravitational Lensing}  \index{1830-211}  \index{Astrochemistry}  
\index{Chemical Abundances}  \index{Isotopes}  \index{Isomers}  
\index{Deuterium} \index{0218+357} \index{Giant Molecular Clouds}
\index{Diffuse Molecular Clouds} \index{Milky Way}
\index{Very Large Array} \index{VLBA} \index{CCH} 
\index{Einstein ring} \index{IRAM interferometer}
\index{Effelsberg} \index{N$_2$H}
\index{CS}
\index{OH}
\index{TMC-1}
\index{L 183}
\index{3C 111}
\index{NRAO 150}
\index{Hyperfine Transitions}
\index{Subthermal Excitation}
\index{Dark Molecular Clouds}
\index{Tausus/Auriga Dark Cloud Complex}
\index{Galactic Center}


\section{Introduction}

The discovery by Wiklind \&\ Combes
of high-redshift molecular absorption toward four radio sources
provides a novel approach of studying the dense interstellar medium of
galaxies at cosmological distances. In two of the detected sources,
1504+377 and 1413+135,
the absorption takes place in the host galaxy of the active galactic
nucleus (AGN) providing the background continuum emission
(Wiklind \&\ Combes 1994, 1996a).
In the other two systems, B0218+357 and PKS~1830$-$211, the absorption occurs
in  galaxies that act as gravitational lenses magnifying the
radio- and millimeter-wavelength emission of a background AGN. In both of
the latter sources the absorbing material,
which is, respectively, at redshifts near 0.685 and 0.89, 
has a remarkably high molecular hydrogen column density exceeding
$10^{22}$ \pscm\ (Wiklind \&\ Combes 1995, 1996b; Menten \&\ Reid 1996).
Particularly toward PKS~1830$-$211, one observes a wide variety
of absorption lines from a series of high dipole moment species that are
well-known constituents of molecular clouds within the Milky Way
(Wiklind \&\ Combes 1996b, 1998).
Here we report interferometric observations targeted at understanding
the spatial distribution and chemical composition of the absorbing clouds
observed toward B0218+357 and PKS~1830$-$211.

\section{Observations}

We have used the VLA to search for absorption from a number of molecules
toward B0218+357 and PKS~1830$-$211. In particular, the $J=1-0$ rotational
ground state lines from the high dipole moment species HCO$^+$, HCN, and
HNC, which are detected toward many Milky Way molecular clouds, have rest
frequencies in the 3 mm band that are redshifted into the 38--49 GHz range
covered by the VLA's ``new'' highest frequency receivers for
$z \approx 0.83$ -- 1.3. At the time of our observations, 13 VLA antennas
were equipped with 38--49 GHz receivers. Toward PKS~1830$-$211, we
detected absorption from the $J = 1-0$ lines of the three molecules mentioned,
as well as from the analogous transitions of their $^{13}$C-substituted
isotopomers. In addition, we detected absorption from the
C$_3$H$_2$, HC$_3$N, H$_2$CO, and C$_2$H molecules.
Our VLA observations allow for accurate optical depth determinations.
Moreover, since some of our observations used the highest resolution
A configuration, we are able to image the PKS~1830$-$211 system at
$0{\rlap.}{''}1$  resolution. The observations described here are
part of an extensive
program of studying molecular absorption toward PKS~1830$-$211 and
B0218+357 that also includes observations with the IRAM interferometer,
the Effelsberg 100 m telescope, and the Very Long Baseline Array (VLBA).

\begin{figure}
  \centerline{
  \psfig{figure=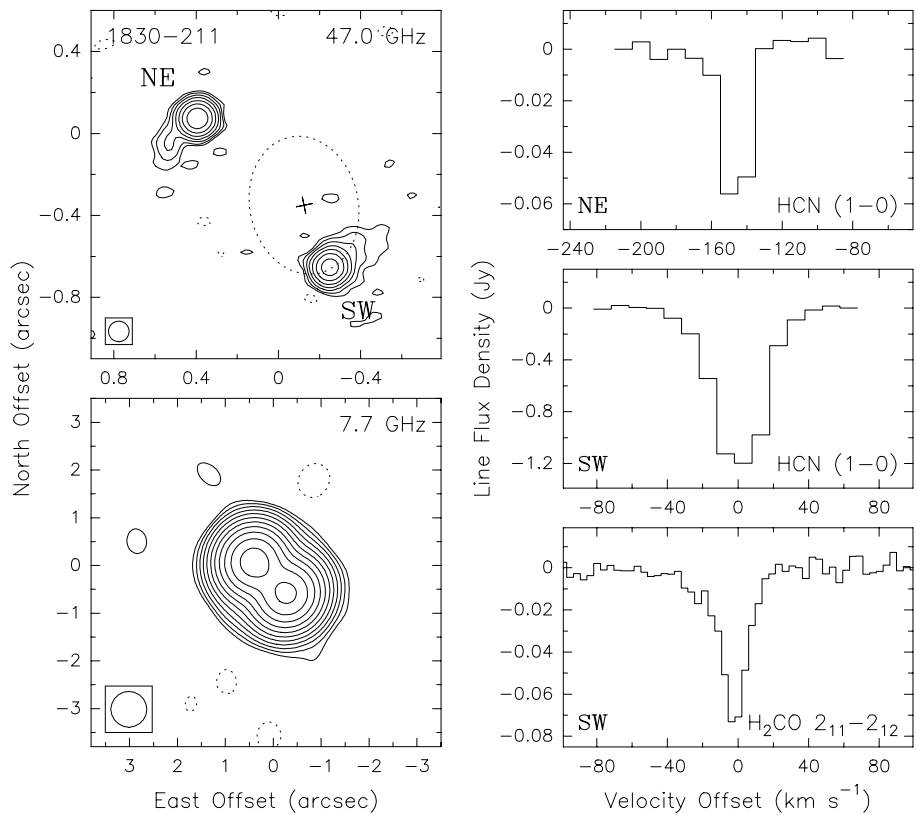,width=12cm}}
  \caption{VLA continuum images and molecular absorption spectra of
    PKS~1830$-$211. The upper and lower left panels show continuum maps
    made with the VLA at frequencies of 47 and 7.7 GHz, respectively.
    Note that the 7.7 GHz map shows a larger field. The 47 and 7.7 GHz
    maps were restored with circular beams, indicated in the lower left map
    corners, of size $0{\rlap.}{''}1$  and $0{\rlap.}{''}75$   (FWHM),
    respectively. The cross and dotted ellipse in the 47 GHz map indicate
    the centroid position, orientation, and core size of the oblate
    spheroidal lens mass distribution assumed in the lens model of 
    Nair et al. 1993, which has a core radius
    of $0{\rlap.}{''}34$ and an
    eccentricity of 0.63. The upper two panels on the right hand side
    show absorption spectra of the HCN $1 - 0$ rotational transition taken
    toward the
    north-eastern (NE) and south-western (SW) images, respectively, while
    the lower right panel shows a spectrum of the $2_{11}-2_{12}$ line of
    H$_2$CO toward the SW image. The velocity scales of all spectra are in
    the heliocentric system with zero velocity corresponding
    to $z = 0.88582$.} \label{fig1830map}
\end{figure}

\section{PKS~1830$-$211: Morphology and Spatial Extent of the Absorption}
The upper left panel of Fig. 1 shows a 47.0 GHz VLA map of
the PKS~1830$-$211 gravitational lens system made with a resolution of
$0{\rlap.}{''}1$. Two compact sources, in the following referred to as the
NE and SW images, with weak extended ``tails'' are seen, displaying the
inversion
symmetry expected 
from gravitational lensing of a core/jet source. With decreasing
frequency, the jet components grow larger (Rao \&\ Subrahmanyan 1988;
Subrahmanyan et al. 1990) and
form an Einstein ring (Jauncey et al. 1991), which is {\it not} resolved in our
$0{\rlap.}{''}75$ resolution 7.7 GHz map shown in the lower left panel of
Fig. 1.

Also shown in Fig. 1 are spectra of the
HCN $J = 1 - 0$ line made toward both images.
Strong absorption with a peak optical depth $\simgreat 2$ is observed
at zero velocity (assuming $z = 0.88582$) toward the compact core of
the SW image, while much weaker absorption is observed in the same transition
toward the NE image at a 147 \kms\ lower velocity. This
general distribution of
the HCN absorption is consistent with the results obtained for the
$2 - 1$ line of this molecule
by Frye, Welch, \&\ Broadhurst (1997) and Wiklind \&\ Combes (1998).
Our higher resolution VLA observations, together with VLBA data on the \hcccn\
$J = 5 - 4$ line (redshifted to 24.1 GHz; Carilli et al. 1998) 
allow for a more detailed analysis of the spatial structure of the
molecular absorption toward PKS~1830$-$211(SW).
As discussed by Carilli et al. (1998), the \hcccn\ absorption is arising from
a region of size larger than 2.5 milliarcseconds
or $10 h^{-1}$ pc,
where $h$ is the Hubble constant in units of 100 \kms~Mpc$^{-1}$ and
$q_0 = 0.5$ is assumed. An upper limit of $\sim 600 h^{-1}$ pc
on the size of the absorbing cloud
follows from the fact that we do not detect HCN 
absorption toward the tail of SW image with an upper limit on the
optical depth of 0.3.

As discussed by Wiklind \&\ Combes (1998), the observed difference in
absorption velocities between the SW and NE images
allows for tight constraints on the lens mass
distribution and, thus, lensing models of PKS~1830$-$211 (see Kochanek \&\
Narayan 1992; Nair, Narasimha, \&\ Rao 1993).
Wiklind \&\ Combes (1998) derive $V_0 \approx 220 \sqrt{D}$ \kms\ for
the rotation velocity, $V_0$, of the $z \approx 0.89$ lensing galaxy.
The ``effective distance'' $D$, measured in Gpc, is equal to
$D_l D_s/D_{ls}$; where $D_l$ is the distance between
the observer and the lens, $D_s$ the distance between the observer and
the source, and $D_{ls}$ the distance between the lens and the source.
All distances are angular diameter distances. Very recently,
a value of 2.507 has been measured for the redshift of the lensed source
in the PKS~1830$-$211 system (see the report by Lidman 1998),
yielding $V_0 \approx 366$ \kms\ and an inclination angle, $i$, of 16$\deg$
($H_0$ = 75 \kms~Mpc$^{-1}$ and $q_0 = 0.5$ are assumed throughout this
paper).  This indicates that the lensing galaxy is a massive early-type
spiral.
There is, however, the
possible complication of multiple lensing, which might be expected, given the
presence of an HI absorption system at a redshift of 0.1927 (Lovell
et al. 1996).

\begin{figure}
  \centerline{
  \psfig{figure=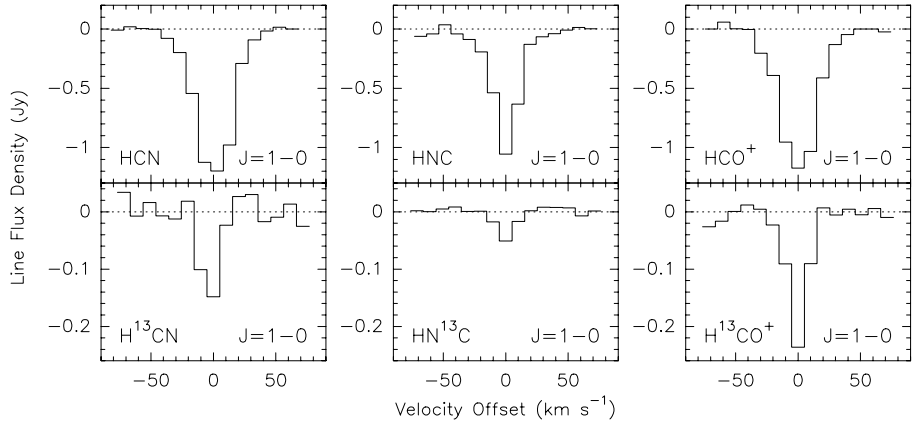,width=12cm}}
  \caption{Spectra of the redshifted $J=1-0$ transition from the main
    isotopomers of HCN, HNC, and HCO$^+$ observed toward PKS~1830$-$211~(SW)
    are
    shown in the upper row, while spectra of the same transition from
   the $^{13}$C substituted isotopomers are shown in the lower row.
   The velocity scales of all spectra are in
    the heliocentric system with zero velocity corresponding
    to $z = 0.88582$.}\label{fig1830iso}
\end{figure}

\begin{figure}
  \centerline{
  \psfig{figure=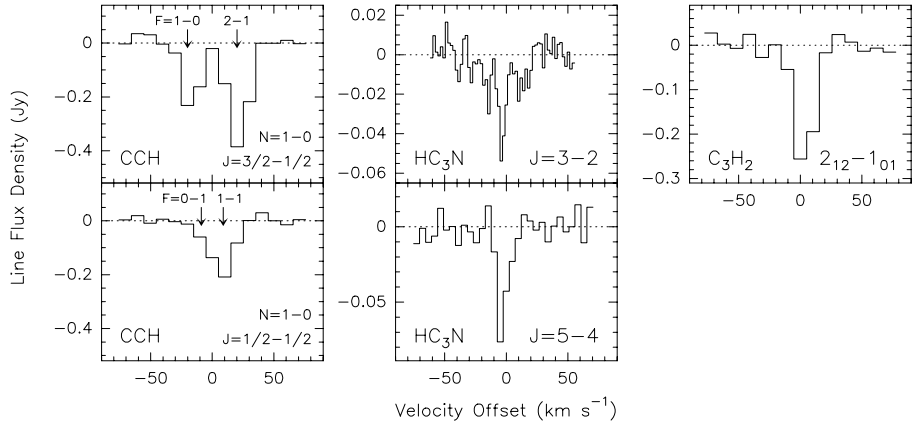,width=12cm}}
  \caption{Redshifted absorption spectra of various hyperfine components
    of the $N=1-0$ transition of CCH (left hand panels), the HC$_3$N
    $J = 3 - 2$ and $5 - 4$ transitions (middle panels), and the $2_{12} -
    1_{01}$ transition of C$_3$H$_2$ (right hand panel) observed with the
    VLA toward PKS~1830$-$211.
       The velocity scales of all spectra are in
    the heliocentric system with zero velocity corresponding
    to $z = 0.88582$.}\label{fig1830cch}
\end{figure}

\def\p  {\phantom{1}}
\def\pp  {\phantom{11}}
\def\ppp  {\phantom{111}}
\def\pppp  {\phantom{1111}}
\def\g  {\phantom{$>$}}
\def\gp  {\phantom{$>$1}}
\def\gpp  {\phantom{$>$11}}
\def\gppp  {\phantom{$>$111}}
\def\gpppp  {\phantom{$>$1111}}

\begin{table}
\caption{Molecules toward PKS1830$-$211 and Galactic Clouds
  \label{tbl-ab}}
\begin{center}\scriptsize
\begin{tabular}{llcccccc}
\tableline
\tableline
\noalign{\vskip 0.06in}
Species &$\mu$& N(PKS~1830$-$211) & \multicolumn{5}{l}{$N$(X)/$N$(CO)$\times 10^6$\dotfill}  \\
~       &(D)& (cm$^{-2}$)
&PKS~1830$-$211
&TMC-1
&L~183
&3C111
&NRAO~150\\
\noalign{\vskip 0.06in}
\tableline
\noalign{\vskip 0.06in}
{\it CO}        &0.11&3.0$\times10^{18}$&   \ppp1&  \gppp1&     \gpp1&  \pp1& \pp1\\
{\it CS}        &1.96&4.8$\times10^{14}$&   \p160&  \gp130&     \gpp9\\
HCO$^+$         &3.30&4.0$\times10^{14}$&   \p130&  \gp110&     \g100&   130&180--1400\\
HCN             &2.98&3.2$\times10^{14}$&   \p110&  \gp130&     \gp40&      &400--2000\\
HNC             &2.70&1.2$\times10^{14}$&   \pp40&  \gp250&   \p$<$50\\
{\it N$_2$H}$^+$&3.40&5.4$\times10^{13}$&   \pp18&  \gpp10&\p$\sim$10\\
CCH             &0.8&7.9$\times10^{14}$&    \p260&\p$>$100&    $<$100\\
H$_2$CO         &2.33&3.8$\times10^{14}$&   \p130&  \gp250&     \g250&   220&590--1200\\
C$_3$H$_2$      &3.4&6.7$\times10^{12}$&    \ppp2&  \gp130&     \gpp3&  \p56&   190\\
HC$_3$N         &3.72&8.1$\times10^{12}$&   \ppp3&  \gpp75&     \gpp1\\ 
{\it OH}        &1.67&4.1$\times10^{15}$&    1400&  \g3800&     \g940&   250\\
\tablenotetext{}{The first, second, and third columns list, respectively, 
  molecular species detected toward PKS~1830$-$211, their
  dipole moments (in Debyes), and column densities in the $z=0.88582$
  absorbing cloud measured toward PKS~1830$-$211(SW).
  Column densities of species listed in italics are {\it not} derived
  from our data but are
  taken from Wiklind \&\ Combes 1996b and, for CO and OH, from
  Gerin et al. 1997 and Chengalur, de Bruyn, \&\ Narasimha
  (this volume), respectively.
  The remaining columns list for
  PKS~1830$-$211 and four well-studied Galactic molecular clouds the 
  column densities measured for the various species multiplied by $10^6$
  and divided by the column density of CO in the clouds in question.
  The Galactic clouds considered are the dark clouds
  TMC-1 and L~183 (also known as L~134~N; see Irvine et al. 1987
  and van Dishoeck et al. 1993 and references therein for abundances)
  and the diffuse clouds
  seen in absorption against
  the extragalactic radio sources 3C111 and NRAO~150 (Liszt \&\ Lucas 1995,
  1998; Lucas \&\ Liszt 1996; and Cox, G{\"u}sten, \&\ Henkel 1988).
  For NRAO~150 the range
  of values observed for the various absorption components occurring at
  different radial velocities is given, except for the case of C$_3$H$_2$.}
\end{tabular}
\end{center}

\end{table}

\section{Astrochemistry at {\it z} = 0.88582}
\subsection{Gas Excitation
  and Temperature of the Cosmic Microwave Background}

Figs. 1, 2, and 3 show examples of absorption line spectra toward
the PKS 1830$-$211 (SW) source resulting from our VLA observations. For some
species we
can compare our measured optical depths with the values determined by
Wiklind \&\ Combes (1996b) for higher excitation lines of the same species.
We find that to within the uncertainties of a few K, the thus determined
rotation temperatures, $T_{\rm rot}$, are consistent with the value of the
cosmic microwave background temperature, $T_{\rm cmb}$, at $z = 0.89$.
All of the species observed by us have high dipole moments (see Table 1).
Even for the relatively low frequency \hcccn\ lines, the critical
densities necessary for thermalization of the level populations
at or near the kinetic
temperature, $T_{\rm kin}$, are $\simgreat 10^4$ cm$^{-3}$. Therefore,
the low rotation temperatures indicate subthermal excitation
due to a relatively low total density of the absorbing cloud
rather than low kinetic temperatures, which for molecular clouds within the
Milky Way are generally greater than 10 K. 

Given that $T_{\rm rot} = T_{\rm cmb}$,
the accurate optical depth determinations afforded by our
interferometer measurements allow for a meaningful estimate of
the cosmic microwave background temperature at $z = 0.89$, for which
big bang theory predicts a value of $(1 + z)~2.73$ K = 5.14 K.
In particular, from the \hcccn\ $J = 3 - 2$ and $5 - 4$
spectra shown in Fig. 3 we derive $T_{\rm rot}$ = $4.5^{+1.5}_{-0.6}$ K.
We note that the errors quoted for the \hcccn\ rotation
temperature are formal uncertainties and do not take into consideration
systematic effects such as variations in the source covering
factor between  14.5 and 24.1 GHz, the frequencies of the redshifted
$3 - 2$ and $5 - 4$ lines.

\subsection{Molecular Abundances -- Comparison with Galactic Clouds}

From our measured optical depths we calculate the total column densities
for \hcop, HCN, HNC, CCH, \hhco, C$_3$H$_2$, and \hcccn\ listed in
Table 1, assuming that the level populations are thermalized at
$T_{\rm cmb}$ = 5.14 K. Table 1 also shows a comparison of the 
abundances of molecules detected toward PKS~1830$-$211 with the
values measured toward four Galactic molecular clouds. Listed are
the abundances 
of the various species (multiplied by $10^6$) relative to the CO column
density. The Galactic clouds in question are the cold, dense dark clouds
TMC-1 and L~183 and the more diffuse clouds seen against the
extragalactic radio sources 3C111 and NRAO~150. The dark clouds have
CO column densities, $N$(CO),  of order a few times $10^{18}$ cm$^{-2}$ and  
[CO/H$_2$] abundance ratios $\approx 8\times10^{-5}$ (Irvine et al. 1985;
van Dishoeck et al. 1993 and references therein); while Liszt \&\ Lucas
(1998) determine
$N$(CO) = $9\times 10^{16}$ cm$^{-2}$ for 3C111 and 
$N$(CO) values in the range (0.25 -- 6.6)$\times 10^{15}$ cm$^{-2}$
for the various radial velocity components of the multiple-cloud
absorption system observed toward NRAO~150. The 3C111 absorber and
most of the absorbing clouds
seen toward NRAO~150 are quite dense compared to 
``typical'' diffuse clouds studied, e.g., by ultraviolet absorption
spectroscopy and  probably have [CO/H$_2$] abundance ratios
exceeding $10^{-5}$. By comparison, for the diffuse cloud sample discussed by
Federman et al. (1994; see also Liszt \&\ Lucas 1998) [CO/H$_2$]
abundance ratios range from a few times $10^{-8}$ to a few times
$10^{-6}$.

It is clear from Table 1 that the PKS~1830$-$211 absorbing cloud has a total
column density and molecular abundances that are quite similar to the values
found in Galactic dark clouds (note that C$_3$H$_2$ and \hcccn\ are
anomalously over-abundant in TMC-1 by Galactic molecular cloud standards).
The $z = 0.88582$ cloud does not show the surprisingly high
abundances of \hcop, HCN, and \hhco\ relative to CO found in the more diffuse
Galactic clouds seen in absorption toward 3C111 and NRAO~150 and other sources,
which are, particularly in the case of \hhco,
a challenge to current models of diffuse cloud chemistry (Turner 1994; 
Liszt \&\ Lucas 1995).

While there are chemical similarities, a comparison of
cloud structure in space and velocity  reveals
remarkable differences between the $z=0.88582$
absorbing cloud and typical Galactic dark clouds such as TMC-1 or L~183.
Assume that for the PKS~1830$-$211 (SW) cloud
the abundance ratios of, say, the widespread CO,
\hcop, and HCN molecules relative to \hh\ 
are similar to the values found for Galactic dark clouds. This
implies a total molecular hydrogen column density of a few times $10^{22}$
\pscm. Assuming further that the molecular hydrogen
density is smaller than $10^{4}$ \pccm,
as implied by the subthermal excitation of all of the high dipole moment
species, we derive a pathlength larger than $\sim 1$ pc along the line
of sight, which is
to be compared with the 13 pc {\it lower} limit of the source size in the
plane of the sky indicated by the VLBA results of Carilli et al. (1998)
discussed above. To compare with Galactic molecular clouds, it is
instructive to take as a reference source the nearby and well-studied
Taurus/Auriga dark cloud complex (see, e.g., Cernicharo 1991), which has
a total extent of $\sim 50$ pc, while individual clouds
within that complex, such as TMC-1, have sizes of one to a few pc. 
The line widths in such clouds are $< 1$ \kms,
and radial velocities are within $\sim 1$ \kms\ of a median value, with
only a few \kms\ variation over the whole extent of the complex.
In contrast, from, e.g., the high quality spectrum of the optically thin
\hhco\ absorption line (see Fig. 1),
we find that the $z=0.88582$ absorption covers a total velocity range of
$\simgreat 35$ \kms\ (FWZP), which is significantly larger than the
velocity spread even of a Galactic giant molecular cloud (GMC), with the
exception of the remarkable GMCs found within $\sim 200$ pc of the
Galactic center. The latter clouds have velocity widths that are
comparable to that of the $z=0.88582$ absorber. We note that compared
to the distance of these GMCs to the Galactic center, the cloud absorbing
PKS~1830$-$211 (SW) is located much further away from the
center of the lensing galaxy,  1.8 kpc in the context of the
model of
Nair et al. (1993) and 3 kpc in the model by Kochanek \&\ Narayan (1992).
This possibly indicates significant differences in molecular cloud
structure and composition between the $z = 0.88582$ lensing galaxy of
PKS~1830$-$211 and the Milky Way.

\begin{figure}
  \centerline{
  \psfig{figure=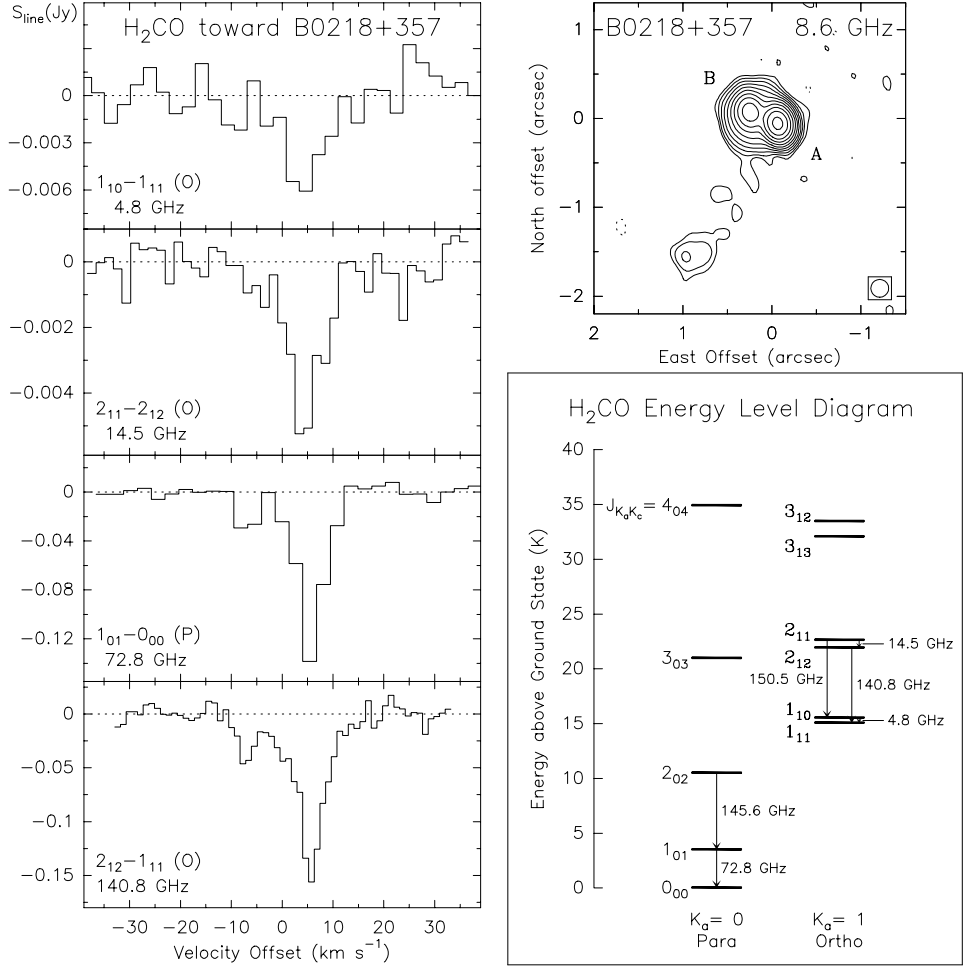,width=12.7cm}}
  \caption{The left four panels show redshifted absorption
    spectra of several \hhco\
     transitions toward B0218+357. The top and bottom spectra were taken
     with the Effelsberg 100 m telescope and the IRAM interferometer,
     respectively, while the other two spectra were observed with the
     VLA. The labels in the left bottom corner of each spectrum list
     the transition and rest frequency. The O or P indicates whether the
     line in question is from ortho- or para-\hhco.
          The velocity scales of all spectra are in
    the heliocentric system with zero velocity corresponding
    to $z = 0.68466$. The upper right panel
     shows a VLA continuum map of the B0218+357 gravitational lens system
     (from  Menten \&\ Reid 1996) made at 8.6 GHz, the frequency of the
     redshifted \hhcotcm\ transition. The \hhco\ absorption occurs to the
     southwestern compact source A. The lower right panel shows an excerpt of
     the \hhco\ rotational energy level diagram. Transitions
     that we measured toward B0218+357 are shown as arrows, with rest
     frequencies indicated.}\label{fighhco}
\end{figure}

\subsection{Isotopic Abundance Ratios}  

Fig. 2 shows spectra of the redshifted $J = 1 - 0$
transitions of HCN, HNC, and \hcop\ together with spectra
of the analogous transitions from the $^{13}$C-substituted isotopomers.
Of the three main isotopic lines, the
HNC line has the lowest optical depth ($\approx 1.2$), affording,
together with the optically thin H$^{13}$CN line, a meaningful
estimate of the [$^{12}$C/$^{13}$C] ratio; we find a value of $\approx 35$.
The  [$^{12}$C/$^{13}$C] ratios determined from HCN and \hcop\ are consistent
with this number. While the [$^{12}$C/$^{13}$C] ratio in the $z = 0.88582$
cloud is smaller than the values found in the solar system and
the local interstellar medium,
[$^{12}$C/$^{13}$C] ratios around 50 are measured for inner Galaxy clouds 
within 4 kpc of the Galactic center, while the Galactic center clouds
themselves have even smaller values around 20 (see
Wilson \&\ Matteucci 1992; Wilson \&\ Rood 1994).

$^{13}$C is only produced in low and intermediate mass stars, while
$^{12}$C is also produced in massive stars. Thus, the [$^{12}$C/$^{13}$C]
ratio is expected to decrease in time and with increasing stellar processing
(see the discussion of Wilson \&\ Matteucci 1992). The relatively low
value measured at $z = 0.88582$ may thus be of interest in the context
of chemical evolution studies.
However, we note that
we only sample one line of sight in the lensing galaxy, which might have
an atypical chemical history.

We also searched unsuccessfully for absorption the DCN $(1-0)$ transition.
Our upper limit and other searches for deuterated molecules toward PKS~1830$-$211
are discussed by Shah et al. (this volume).

\section{A Multi-Transition Study of Formaldehyde at {\it z} = 0.68466
  toward B0218+357}

Formaldehyde (\hhco) was first observed toward the Einstein ring
B0218+357 in its 2 cm rest wavelength \hhcotcm\ $K$-doublet transition
by Menten \&\ Reid (1996), who found absorption toward the
south-western of the two compact images of the background source, labeled
A in Fig. 4
(see also Patnaik et al. 1993 and Patnaik \&\ Porcas, this volume, for
a description of the radio continuum morphology).
The \hhco\ absorption occurs at $z=0.68466$, the same redshift at which
Wiklind \&\ Combes (1995) detect absorption from other molecules.

The populations of the \hhcoscm\ and \hhcotcm\ $K$-doublet transitions
at 6 and 2 cm wavelength, respectively, are sensitively dependent
on the \hh\ density and 
these lines are usually anti-inverted. Statistical equilibrium 
calculations yield  excitation temperatures and optical depths, so that  
comparisons with observations in principle allow
meaningful estimates of the molecular
hydrogen density (see, e.g., Henkel, Walmsley, \&\ Wilson 1980). 
While both the 6 and the 2 cm
transitions have now been detected toward B0218+357
(Fig. 4),
their interpretation is complicated by the fact that
the extent and intensity of the background
continuum emission is strongly dependent
on the observing frequency (see Patnaik \&\ Porcas, this volume).
Modeling of the \hhco\ excitation will be  greatly aided by VLBI observations
and determinations
of the excitation temperatures of the 6 cm and 2 cm lines, which can be
extracted from observations of millimeter-wavelength lines 
interconnecting the energy levels of the centimeter lines (see Fig. 4).
For this
reason, we have also observed a number of millimeter-wavelength
\hhco\ transitions toward \hhco. While spectra for several lines are
shown in Fig. 4, the results of our model calculations will
be presented in a future publication.

Since we have observed both ortho- and para-\hhco\ lines, our
observations allow an estimate of the ortho-to-para ratio, which should
be 3 if the \hhco\ was formed in warm gas (\Tkin $\gg 15$ K) or on hot
dust grain surfaces and $\approx 1$ for formation in cold gas or on
cold grain surfaces.

From a preliminary analysis of the \hhcoparags\ para- and the
$2_{12} - 1_{11}$ ortho-transition (see Fig. 4), we derive an
ortho-to-para \hhco\ ratio of $3.1 \pm 0.2$.
Systematic uncertainties of this value are likely to be small, since,
contrary to the case of the low frequency $K$-doublet transitions
mentioned above, the covering factors of  the two millimeter-wavelength
transitions used in its derivation should be very similar.

\acknowledgments{The National Radio Astronomy Observatory is a 
facility of the National Science Foundation
operated under cooperative agreement by Associated 
Universities, Inc.}


\begin{references}
\reference Carilli, C. L., Menten, K. M., Reid, M. J., Rupen, M. P., \&\
  Claussen, M. 1998, in IAU Colloq. 164: Radio Emission from Galactic and
  Extragalactic Compact Sources, ed. J.~A. Zensus, G.~B. Taylor, \&\ J.~M.
  Wrobel (San Francisco: ASP), 317 
\reference Cernicharo, J. 1991, in The Physics of Star Formation and
Early Stellar Evolution, ed. C. J. Lada \&\ N. D. Kylafis
(Dordrecht: Kluwer), 287
  \reference Cox, P., G{\"u}sten, R, \&\ Henkel, C. 1988, A\&A, 206, 108
\reference Federman, S. R., Strom, C. J., Lambert, D. L., Cardelli, J. A.,
Smith, V. V., Joseph, C. L. 1994, ApJ, 424, 772
  \reference Frye, B., Welch, W. J., \&\ Broadhurst, T. 1997, ApJ, 478, L25
\reference Gerin, M., Phillips, T. G., Benford, D. J., Young, K. H.,
  Menten, K. M., \&\ Frye, B. 1997, ApJ, 488, L31
\reference Henkel, C., Walmsley, C. M., \&\ Wilson, T. L. 1980, A\&A, 82, 41
  \reference Irvine, W. M., Schloerb, F. P., Hjalmarson, \AA., \&\ Herbst, E.,
  1985, in Protostars and Planets II, ed. D. C. Black \&\ M. S. Matthews  
  (Tucson: Univ. Arizona Press), 579
\reference Jauncey, D. L., et al. 1991, Nature, 352, 132
\reference Kochanek, C. S., \&\ Narayan, R. 1992, 401, 461
\reference Lidman, C. 1998, ESO Messenger, 93, 16 
\reference Liszt, H. , \&\ Lucas, 1995, A\&A, 299, 847
\reference Liszt, H. S., \&\ Lucas, 1998, A\&A, 339, 561
\reference Lucas, R., \&\ Liszt, H. 1996, A\&A, 307, 237
\reference Lovell, J. E. J., et al. 1996, ApJ, 472, L5
\reference Menten, K. M., \&\ Reid, M. J. 1996, ApJ, 465, L99
\reference Nair, S., Narasimha, D., \&\ Rao, A. P. 1993, ApJ, 407, 46
\reference Patnaik, A. R., Browne, I. W. A., King, L. J., Muxlow, T. W. B.,
  Walsh, D., \&\ Wilkinson, P. N. 1993, \mnras, 261, 435
\reference Rao, A. P., \&\ Subrahmanyan, R. 1988, MNRAS, 231, 229
\reference Subrahmanyan, R., Narasimha, D., Rao, A. P., \&\ Swarup, G.
1990, MNRAS, 246, 263
\reference Turner, B. E. 1994, ApJ, 437, 658
\reference van Dishoeck, E. F., Blake, G. A., Draine, B. T., \&\ Lunine,
  J. I. 1993, in Protostars \&\ Planets III, ed. E. Levy \&\ J. I. Lunine
  (Tucson: Univ. Arizona Press), 163
\reference Wiklind, T., \&\ Combes, F. 1994,  A\&A, 286, L9
\reference Wiklind, T., \&\ Combes, F. 1995,  A\&A, 299, 382
\reference Wiklind, T., \&\ Combes, F. 1996a, A\&A, 315, 86
\reference Wiklind, T., \&\ Combes, F. 1996b, Nature, 379, 139
\reference Wiklind, T., \&\ Combes, F. 1998,  ApJ, 500, 129
\reference Wilson, T. L., \&\ Rood, R. T. 1994, ARAA, 32, 191
\reference Wilson, T. L., \&\ Matteucci, F. 1992, A\&AR, 4, 1

\end{references}
\end{document}